\magnification=\magstep1
\hfuzz=6pt
\baselineskip=15pt

$ $

\vskip 1in

\centerline{\bf Ultimate physical limits to computation}

\bigskip

\centerline{Seth Lloyd}

\centerline{d'Arbeloff Laboratory for Information Systems and Technology}

\centerline{MIT Department of Mechanical Engineering}

\centerline{MIT 3-160, Cambridge, Mass. 02139}

\centerline{slloyd@mit.edu}

\bigskip

{\bf Computers are physical systems: what they can and cannot do is 
dictated by the laws of physics$^{1-86}$.   In particular, the speed with
which a physical device can process information is limited by
its energy$^{11-26}$ and the amount of information that it can process
is limited by the number of degrees of freedom it possesses$^{5-40}$.
This paper explores the physical limits of computation
as determined by the speed of light $c$, the quantum scale $\hbar$
and the gravitational constant $G$.  As an example, quantitative
bounds are put to the computational power of an `ultimate laptop' 
with a mass of one kilogram confined to a volume of one liter.} 
  
Over the past half century, the amount
of information that computers are capable of processing and the
rate at which they process it has doubled every two years, a
phenomenon known as Moore's law.  A variety of technologies ---
most recently, integrated circuits --- have enabled this exponential
increase in information processing power.  There is no particular
reason why Moore's law should continue to hold: it is a law of
human ingenuity, not of nature.  At some point, Moore's law
will break down.  The question is, When?  Extrapolation of
current exponential improvements over two more decades would
result in computers that process information at the scale of
individual atoms.  Although an Avogadro scale computer that can
act on $10^{23}$ bits might seem implausible, prototype quantum computers 
that store and process information on individual atoms have already been
demonstrated$^{64-65, 76-80}$.  Although existing quantum computers are
small and simple, able to perform a few hundred operations on fewer
than ten quantum bits or `qubits,' the fact that they work at all
indicates that there is nothing in the laws of physics that
forbids the construction of an Avogadro-scale computer.

The purpose of this article is to determine just what limits the
laws of physics place on the power of computers.  At first, this
might seem a futile task: since we don't know the technologies by
which computers one thousand, one hundred, or even ten years in
the future will be constructed, how can we determine the physical
limits of those technologies?  In fact, as will now be shown, a
great deal can be determined concerning the ultimate physical
limits of computation simply from knowledge of the speed of
light, $c= 2.9979 \times 10^{8}$ meters per second, Planck's
reduced constant, $\hbar =  1.0545 \times 10^{-34}$ joule seconds,
and the gravitational constant, $G = 6.673 \times 10^{-11}$ 
meters cubed per kilogram second squared.  Boltzmann's constant,
$k_B = 1.3805 \times 10^{-23}$ joules per degree Kelvin, will also
play a key role in translating between computational
quantitites such as memory space and operations per bit per 
second, and thermodynamic quantitites such as entropy and temperature.
This article reviews previous work on how physics limits the
speed and memory of computers and presents new results: the
derivation of the ultimate speed limit to computation, the 
trade-offs between memory and speed, and the analysis of the
behavior of computers at physical extremes of high temperatures
and densities are novel except as noted.  

Before presenting methods for calculating these limits, it is
important to note that there is no guarantee that these limits will
ever be attained, no matter how ingenious computer designers
become.  Some extreme cases such as the black-hole computer
described below are likely to prove extremely difficult or impossible
to realize.  Human ingenuity has proved great in the past, however,
and before writing off physical limits as unattainable, one should
realize that certain of these limits have already
been attained within a circumscribed context in the construction
of working quantum computers.  The discussion below will note 
obstacles that must be sidestepped or overcome before various
limits can be attained.

\bigskip\noindent{\bf 1. Energy limits speed of computation}

To explore the physical limits of computation, let us calculate
the ultimate computational capacity of a computer with a mass of
one kilogram occupying a volume of one liter, roughly the size of
a conventional laptop.  Such a computer, operating at the limits
of speed and memory space allowed by physics, will be called the 
`ultimate laptop.'

First, ask what limits the laws of physics place on the speed 
of such a device.   As will now be shown, to perform
an elementary logical operation in time $\Delta t$ requires
an average amount of energy 
$ E \geq \pi \hbar/2\Delta t$.
As a consequence, a system with average energy $ E $
can perform a maximum of $2 E / \pi \hbar$
logical operations per second.  A one kilogram computer has
average energy $ E = mc^2 = 8.9874 \times 10^{16}$ joules. 
Accordingly, the ultimate laptop can perform a
maximimum of $ 5.4258 \times 10^{50}$ operations per second.

\vfill\eject
\bigskip\noindent{\bf 1.1 Maximum speed per logical operation}

For the sake of convenience, the ultimate laptop will be taken 
to be a digital computer.  Computers that operate on 
non-binary or continuous variables obey similar limits to those
that will be derived here.  A digital computer performs computation by 
representing information in the terms of binary digits or bits, 
which can take the value $0$ or $1$, and then processes that information
by performing simple logical operations such as $AND$, $NOT$ 
and $FANOUT$.  The actual physical
device that performs a logical operation is called a logic gate. 
The operation, $AND$, for instance, takes two binary 
inputs $X$ and $Y$ and returns the output 1 if and only if both $X$ 
and $Y$ are 1; otherwise it returns the output 0.  Similarly, $NOT$ 
takes a single binary input $X$ and returns the output 1 if $X=0$ 
and 0 if $X=1$.  $FANOUT$ takes a single binary input $X$ and returns two
binary outputs, each equal to $X$.  Any boolean function can be
constructed by repeated application of $AND$, $NOT$ and $FANOUT$.
A set of operations that allows the construction of arbitrary
boolean functions is called universal.  The actual physical
device that performs a logical operation is called a logic gate.

How fast can a digital computer perform a logical operation? 
During such an operation, the bits in the computer on which the
operation is performed go from one state to another.
The problem of how much energy is required for information processing
was first investigated in the context of communications theory
by Levitin$^{11-16}$, Bremermann$^{17-19}$, Beckenstein$^{20-22}$
and others, who showed that the laws of quantum mechanics
determine the maximum rate at which a system with spread in
energy $\Delta E$ can move from one distinguishable state to
another.  In particular, the correct interpretation of the
time-energy Heisenberg uncertainty principle 
$\Delta E \Delta t \geq \hbar$ is not that it 
takes time $\Delta t$ to measure energy to an accuracy
$\Delta E$ (a fallacy that was put to rest by Aharonov
and Bohm$^{23-24}$) but rather that
that a quantum state with spread in energy $\Delta E$
takes time at least $\Delta t = \pi \hbar/2\Delta E$ to evolve
to an orthogonal (and hence distinguishable) state$^{23-26}$.  
More recently, Margolus and Levitin$^{15-16}$ extended this result to
show that a quantum system with average energy $ E $
takes time at least $\Delta t = \pi \hbar/2 E$ to evolve
to an orthogonal state.

\bigskip\noindent{\bf 1.2 Performing quantum logic operations}

As an example, consider the operation $NOT$ performed on a quantum
bit or `qubit' with logical states $|0\rangle$ and $|1\rangle$.  
(For readers unfamiliar with quantum mechanics,
the `bracket' notation $|~\rangle$ signifies that whatever is
contained in the bracket is a quantum-mechanical variable;
$|0\rangle$ and $|1\rangle$ are vectors in a two-dimensional
vector space over the complex numbers.)  
To flip the qubit, one can apply a potential 
$H = E_0|E_0\rangle \langle E_0| + E_1 |E_1\rangle \langle E_1|$ 
with energy eigenstates 
$|E_0\rangle = (1/\sqrt 2)(|0\rangle + |1\rangle)$ and 
$|E_1\rangle = (1/\sqrt 2)(|0\rangle - |1\rangle)$. 
Since $|0\rangle = (1/\sqrt 2)(|E_0\rangle + |E_1\rangle)$ and
$|1\rangle = (1/\sqrt 2)(|E_0\rangle - |E_1\rangle)$,
each logical state $|0\rangle$, $|1\rangle$ has spread in
energy $\Delta E = (E_1-E_0)/2$.  It is easy to verify that
after a length of time $\Delta t = \pi \hbar/2 \Delta E$
the qubit evolves so that $|0\rangle \rightarrow |1\rangle$
and $|1\rangle \rightarrow |0\rangle$.  That is, applying the
potential effects a $NOT$ operation in a time that attains the
limit given by quantum mechanics.  Note that the average energy $E$
of the qubit in course of the logical operation is 
$ \langle 0 | H| 0\rangle = \langle 1| H|1\rangle = (E_0 + E_1)/2 
= E_0 + \Delta E$.  Taking the ground-state energy $E_0 = 0$
gives $ E = \Delta E$.  So the amount of time
it takes to perform a $NOT$ can also be written
$\Delta t = \pi \hbar/2 E $.   It is straightforward
to show$^{15-16}$ that no quantum system with average energy 
$E$ can move to an orthogonal state in a time 
less than $\Delta t$.  That is, the speed with which a logical
operation can be performed is limited not only by the spread in
energy, but by the average energy.  This result will prove to be
a key component in deriving the speed limit for the ultimate laptop.

$AND$ and $FANOUT$ can be enacted in a way that is analogous
to the $NOT$ operation.  A simple way to
perform these operations in a quantum-mechanical context is
to enact a so-called Toffoli or Controlled-Controlled-$NOT$
operation$^{31}$.  This operation takes three binary inputs, $X$, $Y$, and
$Z$ and returns three outputs, $X'$, $Y'$, and $Z'$.
The first two inputs pass through unchanged: $X'=X$,
$Y'=Y$.  The third input passes through unchanged unless
both $X$ and $Y$ are 1, in which case it is flipped.
This is universal in the sense that suitable choices of
inputs allows the construction of $AND$, $NOT$, and
$FANOUT$.  When the third input is set to zero, $Z=0$,
then the third output is the $AND$ of the first two:
$Z'= X~AND~Y$.  So $AND$ can be constructed.  When the
first two inputs are 1, $X=Y=1$, the third output is
the $NOT$ of the third input, $Z' = NOT~Z$.  Finally,
when the second input is set to 1, $Y=1$, and the third
to zero, $Z=0$, the first and third output are the $FANOUT$
of the first input, $X'=X$, $Z'=X$.  So arbitrary boolean
functions can be constructed from the Toffoli operation
alone. 

By embedding a Controlled-Controlled-$NOT$ gate in a quantum
context, it is straightforward to see that $AND$ and $FANOUT$,
like $NOT$, can be performed 
at a rate $ 2 E/ \pi \hbar$ times per second,
where  $ E$ is the average energy of the
logic gate that performs the operation.  More complicated logic
operations that cycle through a larger number of quantum states
(such as those on non-binary or continuous quantum variables)
can be performed at a rate $  E/ \pi \hbar$  ---
half as fast as the simpler operations$^{15-16}$.  Existing quantum logic
gates in optical-atomic and NMR quantum computers actually attain
this limit.  In the case of $NOT$, $ E$ is the 
average energy of interaction of the qubit's dipole moment
(electric dipole for optic-atomic qubits and nuclear magnetic dipole
for NMR qubits) with the applied electromagnetic field.
In the case of multi-qubit operations such as the Toffoli
or simpler two bit Controlled-$NOT$ operation, which flips the 
second bit if and only if the first bit is 1, $ E$
is the average energy in the interaction between the physical systems that
register the qubits.

\bigskip\noindent{\bf 1.3 Ultimate limits to speed of computation}

We are now in a position to derive our first physical limit to
computation: energy limits speed.
Suppose that one has a certain amount of energy $E$ to allocate
to the logic gates of a computer.  The more energy one allocates
to a gate, the faster it can perform a logic operation.  The total
number of logic operations performed per second is equal to
the sum over all logic gates of the operations per second
per gate.  That is, a computer can perform no more than
$$\sum_\ell 1/\Delta t_\ell \leq 
\sum_\ell 2 E_\ell/\pi\hbar = 
 2 E/\pi\hbar\eqno(1)$$
operations per second.  In other words, the rate at which a computer 
can compute is limited by its energy.  (Similar limits have been
proposed by Bremmerman in the context of the minimum energy required 
to communicate a bit$^{17-19}$.  These limits have been criticized, however, 
for misinterpreting the energy-time uncertainty relation$^{21}$, and for
failing to take into account the role of degeneracy of energy 
eigenvalues$^{13-14}$ and the role of nonlinearity in communications$^{7-9}$.)  
Applying this result to a one kilogram computer with energy 
$ E= mc^2 = 8.9874 \times 10^{16}$ joules show that our ultimate
laptop can perform a maximum of $5.4258  \times 10^{50}$ operations 
per second.  

\bigskip\noindent{\bf 1.4 Parallel and serial operation}

An interesting feature of this limit is that it is independent of
computer architecture.  One might have thought that a computer
could be sped up by parallelization, i.e., by taking the energy
and dividing it up amongst a large number of subsystems computing
in parallel.  This is not the case: if one spreads the energy
$ E$ amongst $N$ logic gates, each one
operates at a rate $2E/\pi\hbar N$. 
The total number of operations per second, $N2E/\pi\hbar N =
2E/\pi\hbar$, remains 
the same.  If the energy is allocated to fewer logic gates
(more serial operation), the rate $1/\Delta t_\ell$ 
at which they operate and the spread in energy per gate 
$\Delta E_\ell$ go up.  If the energy is allocated to
more logic gates (more parallel operation) then the rate
at which they operate and the spread in energy per gate go down.
Note that in this parallel case, the overall
spread in energy of the computer as a whole is considerably smaller 
than the average energy: in general $\Delta E =
\sqrt{ \sum_\ell \Delta E_\ell^2} \approx \sqrt N \Delta E_\ell$
while $ E  = \sum E_\ell 
\approx N E_\ell $.
Parallelization can help perform certain computations more
efficiently, but it does not alter the total number of
operations per second.  As will be seen below, the degree of parallelizability
of the computation to be performed determines the most efficient 
distribution of energy among the parts of the computer.  Computers
in which energy is relatively evenly distributed over a larger
volume are better suited for performing parallel computations.
More compact computers and computers with an uneven distribution
of energy are better for performing serial computations.

\bigskip\noindent{\bf 1.5 Comparison with existing computers}

Conventional laptops operate much more slowly than the ultimate
laptop.  There are two reasons for this inefficiency.  First,
most of the energy is locked up in the mass of the particles of
which the computer is constructed, leaving only an infinitesimal
fraction for performing logic.  Second, a conventional computer
employs many degrees of freedom (billions and billions of
electrons) for registering a single bit.  From the physical
perspective, such a computer operates in a highly redundant
fashion.  There are good technological reasons for such
redundancy: conventional designs rely on redundancy for
reliability and manufacturability.  In the present discussion,
however, the subject is not what computers are but what they
might be.  The laws of physics do not require redundancy to
perform logical operations:
recently constructed quantum microcomputers use one quantum
degree of freedom for each bit and operate at the Heisenberg limit
$\Delta t = \pi \hbar/ 2\Delta E$ for the time needed to flip a 
bit$^{64-65, 76-80}$.  Redundancy is required for error correction,
however, as will be discussed below.

In sum, quantum mechanics provides a simple answer to the
question of how fast information can be processed using a given
amount of energy.  Now it will be shown that thermodynamics and 
statistical mechanics provide a fundamental limit to how many bits of
information can be processed using a given amount of energy
confined to a given volume.  Available energy necessarily limits
the rate at which computer can process information.  
Similarly, the maximum entropy of a physical system
determines the amount of information it can process.    
Energy limits speed.  Entropy limits memory.

\bigskip\noindent{\bf 2. Entropy limits memory space}

The amount of information that a physical system can store and 
process is related to the number of distinct physical states 
accessible to the system.  A collection of $m$ two-state
systems has $2^m$ accessible states and can register $m$
bits of information.  In general, a system with $N$ accessible states
can register $\log_2 N$ bits of information.  But it has been
known for more than a century that the number of accessible
states of a physical system, $W$, is related to its 
thermodynamic entropy by the formula: $S = k_B \ln W$,
where $k_B$ is Boltzmann's constant$^{}$.
(Although this formula is inscribed on Boltzmann's tomb, it is originally
due to Planck: before the turn of the century, $k_B$ was often known
as Planck's constant.)  

The amount of information that can be registered by a physical 
system is $I=S( E )/k_B\ln 2$, where $S( E )$
is the thermodynamic entropy of a system with expectation value for
the energy $ E$.  Combining this formula with the formula
$2 E/\pi\hbar$ for the number of logical operations that
can be performed per second, we see that when it is 
using all its memory, the number of operations per
bit per second that our ultimate laptop can perform is
$k_B 2\ln 2 E  /\pi\hbar S \propto k_B T/\hbar$,
where $T = (\partial S/\partial E)^{-1}$ is 
the temperature of a kilogram of matter in a maximum entropy
in a liter volume.
The entropy governs the amount of information the system can
register and the temperature governs the number of operations
per bit per second it can perform.

Since thermodynamic entropy effectively
counts the number of bits available to a physical system,
the following derivation of the memory space available to
the ultimate laptop is based on a thermodynamic treatment
of a kilogram of matter confined to a liter volume, in a
maximum entropy state.  Throughout this derivation, it is
important to keep in mind that although the memory space
available to the computer is given by the entropy of
its thermal equilibrium state, the {\it actual} state of the 
ultimate laptop as it performs a computation is completely
determined, so that its entropy remains always equal to zero.   
As above, we assume that we have complete control over the
actual state of the ultimate laptop, and are able to guide
it through its logical steps while insulating it from all
uncontrolled degrees of freedom.  As the following discussion
will make clear, such complete control will be difficult
to attain.

\bigskip\noindent{\bf 2.1 Entropy, energy, and temperature}

In order to calculate the number of operations per second that
could be performed by our ultimate laptop, we assumed
that the expectation value of the energy was $ E$.  
Accordingly, the total number of bits of memory space available to the
computer is $S(E,V)/k_B \ln 2$, where $S(E,V)$
is the thermodynamic entropy of a system with expectation value of the
energy $E$ confined to volume $V$.  The entropy of
a closed system is normally given by the so-called microcanonical
ensemble, which fixes both the average energy {\it and} the spread
in energy $\Delta E$, and assigns equal probability to all states
of the system within a range $[E,E+\Delta E]$.  In the case of the
ultimate laptop, however, we wish to fix only the average energy,
while letting the spread in energy vary according to whether the
computer is to be more serial (fewer, faster gates, larger spread
in energy) or parallel (more, slower gates, smaller spread in
energy).  Accordingly, the ensemble that should be used to calculate
the thermodynamic entropy and the memory space available is the
canonical ensemble, which maximizes $S$ for fixed average energy 
with no constraint on the spread in energy $\Delta E$.  
The canonical ensemble tells how many bits of memory are available
for all possible ways of programming the computer while
keeping its average energy equal to $E$.  In any given
computation with average energy $E$ the ultimate laptop will be  
in a pure state with some fixed spread of energy, and will
explore only a small fraction of its memory space.

In the canonical ensemble a state with energy $E_i$ has probability
$p_i=(1/Z(T))e^{-E_i/k_B T}$ where  $Z(T) = \sum_i e^{-
E_i/k_B T}$ is the partition function, and the temperature $T$ is chosen
so that $E=\sum_i p_i E_i$.  The entropy is 
$S= -k_B \sum_i p_i \ln p_i = E/T + k_B\ln Z$.  
The number of bits of memory space available to the 
computer is $S/k_B\ln 2$.  
The difference between the entropy as calculated using the canonical
ensemble and that calculated using the microcanonical ensemble
is minimal.  There is some subtlety involved in using the canonical ensemble
rather than the more traditional microcanonical ensemble, however.
The canonical ensemble is normally used
for open systems that interact with a thermal bath at temperature $T$.
In the case of the ultimate laptop, however, it is applied to a
{\it closed} system to find the maximum entropy given a fixed expectation
value for the energy.  As a result, 
the temperature $T= (\partial S / \partial E)^{-1}$ plays
a somewhat different role in the context of physical limits of 
computation than it does in the case of an ordinary thermodynamic
system interacting with a thermal bath.  Integrating the relationship
$T= (\partial S / \partial E)^{-1}$ over $E$ yields
$T=CE/S$, where $C$ is a constant of order unity (e.g.,
$C=4/3$ for black-body radiation, $C=3/2$ for an ideal gas,
and $C=1/2$ for a black hole).  Accordingly, the temperature
governs the number of operations per bit per second, $k_B\ln 2
E/\hbar S \approx k_B T/\hbar$, that a system can perform.  
As will become clear, the relationship between 
temperature and operations per bit per second is useful in investigating
computation under extreme physical conditions.

\bigskip\noindent{\bf (Box: The role of thermodynamics in computation}.  
The fact that entropy and information  
are intimately linked has been known since Maxwell introduced his
famous `demon' well over a century ago$^{1}$.  Maxwell's demon is an 
hypothetical being that uses its information-processing ability 
to reduce the entropy of a gas.  The first results in physics 
of information processing were derived in attempts to understand
how Maxwell's demon could function$^{1-4}$.  The role of thermodynamics
in computation has been repeatedly examined over the last 
half century.  In the 1950's, von Neumann$^{10}$ speculated that each logical 
operation performed in a computer at temperature $T$ must dissipate energy
$k_B T \ln 2$, thereby increasing entropy by $k_B \ln 2$.
This speculation proved to be false.  The precise,
correct statement of the role of entropy in computation was due
to Landauer$^{5}$, who showed that reversible, i.e. one-to-one, logical
operations such as $NOT$ can be performed without dissipation in
principle, but that irreversible, many-to-one operations such as
$AND$ or $ERASE$ require dissipation at least $k_B \ln 2$ for each bit
of information lost.  ($ERASE$ is a one-bit logical operation that
takes a bit, 0 or 1, and restores it to 0.)
The argument behind Landauer's principle
can be readily understood$^{37}$.  Essentially, the
one-to-one dynamics of Hamiltonian systems implies that when a
bit is erased the information that it contains has to go
somewhere.  If the information goes into observable degrees of
freedom of the computer, such as another bit, then it has not
been erased but merely moved; but if it goes into unobservable
degrees of freedom such as the microscopic motion of molecules
it results in an increase of entropy of at least $k_B\ln 2$.

In 1973, Bennett$^{28-30}$ showed that all computations could be performed
using reversible logical operations only.  Consequently, by
Landauer's principle, computation does not require
dissipation.  (Earlier work by Lecerf$^{27}$ had anticipated the
possibility of reversible computation, but not its physical
implications.  Reversible computation was discovered
independently by Fredkin and Toffoli$^{31}$.)  
The energy used to perform a logical operation
can be `borrowed' from a store of free energy such as a battery,
`invested' in the logic gate that performs the operation, and
returned to storage after the operation has been performed,
with a net `profit' in the form of processed information.  
Electronic circuits based on reversible logic have been built
and exhibit considerable reductions in dissipation 
over conventional reversible circuits$^{33-35}$.

Under many circumstances it may prove useful to perform 
irreversible operations such as erasure.  If our computer is subject
to an error rate of $\epsilon$ bits per second, for example, then
error-correcting codes can be used to detect those errors and
reject them to the environment at a dissipative cost of 
$\epsilon k_B T_E \ln 2$ joules per second, where $T_E$ is
the temperature of the environment.  ( $k_B T \ln 2$ is the
minimal amount of energy required to send a bit down an information
channel with noise temperature $T$.$^{14}$)  Such error-correcting
routines in our ultimate computer function as working analogs
of Maxwell's demon, getting information and using it to reduce
entropy at an exchange rate of $k_B T \ln 2$ joules per bit.
In principle, computation
does not require dissipation.  In practice, however, any
computer -- even our ultimate laptop -- will dissipate energy.

The ultimate laptop must reject errors to the environment at a
high rate to maintain reliable operation.  To estimate the rate
at which it can reject errors to the environment, assume that 
the computer encodes erroneous bits in the form of black-body
radiation at the characteristic temperature
$5.87 \times 10^8$ K of the computer's memory.$^{21}$ 
The Stefan-Boltzmann law for black-body radiation then implies
that the number of bits per unit area than can be sent out
to the environment is 
${\cal B}= \pi^2 k_B^3 T^3/60\ln2 \hbar^3 c^2 = 7.195 \times
10^{42}$ bits per meter$^2$ per second.  Since the ultimate laptop
has a surface area of $10^{-2}$ square meters and is performing
$\approx 10^{50}$ operations per second, it must have an
error rate of less than $10^{-10}$ per operation in order
to avoid over-heating.  Even if it achieves such an error rate,
it must have an energy throughput (free energy in and thermal
energy out) of $4.04 \times 10^{26}$ watts --- turning over its
own rest mass energy of $mc^2\approx 10^{17}$ joules in a nanosecond!
The thermal load of correcting large
numbers of errors clearly suggests the necessity of operating
at a slower speed than the maximum allowed by the laws of physics.
{\bf )}   

\bigskip\noindent{\bf 2.2 Calculating the maximum memory space}

To calculate exactly the maximum entropy for a kilogram of matter in
a liter volume would require complete knowledge of the dynamics of 
elementary particles, quantum gravity, etc.  We do not possess
such knowledge.  However, the maximum entropy can readily be 
estimated by a method reminiscent of that used to calculate 
thermodynamic quantities in the early universe$^{87}$.  The idea
is simple: model the volume occupied by the computer as a collection
of modes of elementary particles with total average
energy $E$.  The maximum entropy is obtained by
calculating the canonical ensemble over the modes.  Here, we
supply a simple derivation of the maximum memory space available
to the ultimate laptop.  A more detailed discussion of how to 
calculate the maximum amount of information that can be stored
in a physical system can be found in the work of Bekenstein$^{19-21}$. 

For this calculation, assume that
the only conserved quantities other than the computer's energy are
angular momentum and electric charge, which we take to be zero.
(One might also ask that baryon number be conserved, but as will
be seen below, one of the processes that could take place within
the computer is black hole formation and evaporation, which does
not conserve baryon number.)  At a particular 
temperature $T$, the entropy is dominated by the contributions
from particles with mass less than $k_B T/2c^2$.  The $\ell$'th such 
species of particle contributes energy $E = r_\ell \pi^2 V (k_B T)^4/
30 \hbar^3 c^3$ and entropy 
$S = 2 r_\ell k_B\pi^2 V (k_B T)^3/45 \hbar^3 c^3 = 4E/3T$ 
where $r_\ell$ is equal to the number of particles/antiparticles in
the species (i.e., 1 for photons, 2 for electrons/positrons) times
the number of polarizations (2 for photons, 2 for electrons/positrons)
times a factor that reflects particle statistics (1 for bosons,
7/8 for fermions)$^{}$.   As the formula for $S$ in terms of $T$ shows, 
each species contributes $(2\pi)^5 r_\ell /90 \ln 2 \approx 10^{2}$ 
bits of memory 
space per cubic thermal wavelength $\lambda_T^3$ where
$\lambda_T = 2\pi \hbar c/k_B T$.  
Re-expressing the formula for entropy as a 
function of energy, our estimate for the maximum entropy is 
$$S = (4/3) k_B (\pi^2 r V/ 30 \hbar^3 c^3)^{1/4} E^{3/4}
=k_B\ln 2 I,\eqno(2)$$ 
where
$r=\sum_\ell r_\ell$.  Note that $S$ depends only insensitively 
on the total number of species with mass less than $k_B T/2c^2$.

A lower bound on the entropy can be obtained by assuming
that energy and entropy are dominated by black body radiation
consisting of photons.
In this case, $r=2$, and for a one kilogram computer confined
to a volume of a liter we have $k_BT = 8.10 \times 10^{-15}$
joules, or $T = 5.87 \times 10^8$ K.  The entropy is $S = 2.04
\times 10^{8}$ joule/K, which corresponds to an amount of
available memory space $I = S/k_B \ln 2 = 2.13 \times 10^{31}$
bits.  When the ultimate laptop is using all its memory space
it can perform $2\ln 2 k_B E/\pi \hbar S =
3\ln 2 k_B T/ 2\pi \hbar \approx 10^{19}$ operations per bit
per second.  As the number of operations per second $2E/\pi\hbar$
is independent of the number of bits available, the number
of operations per bit per second can be increased by using
a smaller number of bits.  In keeping with the prescription
that the ultimate laptop operates at the absolute limits given
by physics, in what follows, we assume that all available
bits are used.

This estimate for the maximum entropy could be improved (and
slightly increased) by
adding more species of massless particles (neutrinos and
gravitons) and by taking into effect the presence of electrons
and positrons.  Note that $k_B T/ 2 c^2 = 4.51 \times 10^{-32}$
kilograms, compared with the electron mass of $9.1 \times 10^{-31}$
kilograms.  That is, our kilogram computer in a liter is close
to a phase transition at which electrons and positrons are
produced thermally.  A more exact estimate of the maximum entropy
and hence the available memory space would be straightforward
to perform, but the details of such a calculation would
detract from our general exposition, and could only serve
to alter $S$ slightly.  $S$ depends insensitively on the
number of species of effectively massless particles: a change
of $r$ by a factor of $10,000$ serves only to increase $S$ by
a factor of $10$.

\bigskip\noindent{\bf 2.3 Comparison with current computers}

The amount of information that can be stored by the ultimate
laptop, $\approx 10^{31}$ bits, is much higher than the 
$\approx 10^{10}$ bits stored on current laptops.  This is
because conventional laptops use many degrees of freedom to
store a bit where the ultimate laptop uses just one.
There are considerable advantages to using many degrees of
freedom to store information, stability and controllability 
being perhaps the most important.  Indeed, as the above calculation
indicates, in order to take full advantage of the memory space
available, the ultimate laptop must turn all its matter into
energy.  A typical state of the ultimate laptop's memory
looks like a plasma at a billion degrees Kelvin:
the laptop's memory looks like a thermonuclear explosion
or a little piece of the Big Bang!  Clearly, packaging
issues alone make it unlikely that this limit can be obtained,
even setting aside the difficulties of stability and control. 

Even though the ultimate physical limit to how much information
can be stored in a kilogram of matter in a liter volume is
unlikely to be attained, it may nonetheless be possible to
get a fair way along the road to such bit densities.
In other words, the ultimate limits to memory space may prove
easier to approach than the ultimate limits to speed.
Following Moore's law, the density of bits in a computer has gone down from
approximately one per cm$^2$ fifty years ago to one per $\mu{\rm m}^2$
today, an improvement of a factor of $10^8$.  It is not inconceivable
that a similar improvement is possible over the course of the next
fifty years.  In particular, there is no physical reason why it should
not be possible to store one bit of information per atom.
Indeed, existing NMR and ion-trap quantum computers already
store information on individual nuclei and atoms (typically in
the states of individual nuclear spins or in hyperfine atomic
states).  Solid-state NMR with high gradient fields or quantum
optical techniques such as spectral hole-burning provide 
potential technologies for storing large quantities of information
at the atomic scale.   A kilogram of ordinary matter holds
on the order of $10^{25}$ nuclei.  If a substantial fraction
of these nuclei can be made to register a bit, then one can
get quite close to the ultimate physical limit of memory without having
to resort to thermonuclear explosions.  If, in addition, one
uses the natural electromagnetic interactions between nuclei
and electrons in the matter to perform logical operations,
one is limited to a rate of approximately $10^{15}$ operations
per bit per second, yielding an overall information processing
rate of $ \approx 10^{40}$ operations per second in ordinary matter.
Although less than the $\approx 10^{51}$ operations per second in
the ultimate laptop, the maximum information processing rate 
in `ordinary matter' is still quite respectable.
Of course, even though such an `ordinary matter' ultimate
computer need not operate at nuclear
energy levels, other problems remain: for example, the high number
of bits still suggests substantial input/output problems.
At an input/output rate of $10^{12}$ bits per second, 
an Avogadro-scale computer with $10^{23}$ bits 
would take about $10,000$ years to perform a serial read/write 
operation on the entire memory.  Higher throughput and parallel
input/output schemes are clearly required to take advantage
of the entire memory space that physics makes available.

\bigskip\noindent{\bf 3. Size limits parallelization}

Up until this point, we have assumed that our computer occupies
a volume of a liter.  The previous discussion, however, indicates
that benefits are to be obtained by varying the volume to which
the computer is confined.
Generally speaking, if the computation to be performed is highly
parallelizable or requires many bits of memory, the volume of the
computer should be greater and the energy available
to perform the computation should be spread out evenly amongst the
different parts of the computer.  Conversely, if the computation to
be performed is highly serial and requires fewer bits of memory,
the energy should be concentrated in particular parts of the computer.

A good measure of the degree of parallelization in a computer
is the ratio between time it takes to communicate from one side
of the computer to the other, and the average time it takes to 
perform a logical operation. 
The amount of time it takes to send a message from one side
of a computer of radius $R$ to the other is $t_{\rm com}= 2R/c$. 
The average time it takes a bit to
flip in the ultimate laptop is the inverse of the number of 
operations per bit per second calculated above: 
$t_{\rm flip} = \pi\hbar S / k_B 2\ln 2 E $.
Our measure of the degree of parallelization in the ultimate
laptop is then 
$$t_{\rm com}/t_{\rm flip} = 
k_B 4\ln 2R E / \pi\hbar c S \propto k_B RT/\hbar c = 2\pi R/\lambda_T.
\eqno(3)$$   
That is, the amount of time it takes to communicate from 
one side of the computer to the other, divided by the amount
of time it takes to flip a bit, is approximately equal to the
ratio between the size of the system and its thermal wavelength.
For the ultimate laptop, with $2R=10^{-1}$ meters, $2E/\pi\hbar
\approx 10^{51}$ operations per second, and $S/k_B\ln 2 \approx
10^{31}$ bits, $t_{\rm com}/t_{\rm flip} \approx 10^{10}$.
The ultimate laptop is highly parallel.
A greater degree of serial computation can be obtained at
the cost of decreasing memory space by 
compressing the size of the computer or making the distribution
of energy more uneven.  As ordinary matter obeys
the Beckenstein bound$^{20-22}$, $k_BR E/\hbar c S >1/2\pi$, however, 
as one compresses the computer $t_{\rm com}/t_{\rm flip}
\approx k_B R E/\hbar c S $ will
remain greater than one: i.e., the operation will still
be somewhat parallel.  Only at the ultimate limit of compression ---
a black hole --- is the computation entirely serial.

\bigskip\noindent{\bf 3.1 Compressing the computer allows more
serial computation}

Suppose that one wants to perform a highly serial computation 
on few bits.  Then it is advantageous to compress the size of the
computer so that it takes less time to send signals from one side
of the computer to the other at the speed of light.  
As the computer gets smaller, keeping the energy fixed,
the energy density inside the computer goes up. 
As the energy density in the computer goes up, different regimes
in high energy physics are necessarily explored in the course of
the computation.  First the weak unification scale is reached,
then the grand unification scale.  Finally, as the linear size of
the computer approaches its Schwarzchild radius, the Planck scale
is reached.  (No known technology could possibly achieve such
compression.)  At the Planck scale, gravitational effects and
quantum effects are both important: 
the Compton wavelength of a particle of mass $m$, $\lambda_C =
2\pi\hbar/mc$ is on the order of its Schwarzschild radius, $2Gm/c^2$.
In other words, to describe behavior at  
length scales of the size $\ell_P = \sqrt{\hbar G/ c^3}=
1.616 \times 10^{-35}$ meter, time scales $t_P =
\sqrt{\hbar G/c^5}= 5.391 \times 10^{-44}$ second, and
mass scales of $m_P = \sqrt{\hbar c/G} = 2.177 \times
10^{-8}$ kilograms, a unified
theory of quantum gravity is required.   We do not currently
possess such a theory.  Nonetheless, although we do not know the 
exact number of bits that can be registered by a one kilogram computer 
confined to a volume of a liter, we do know the exact number of bits that
can be registered by a one kilogram computer that has been compressed
to the size of a black hole$^{90}$.  This is because the entropy
of a black hole has a well-defined value.  

In the following discussion,
we use the properties of black holes to place limits on the
speed, memory space, and degree of serial computation that could
be approached by compressing a computer to the smallest possible size.
Whether or not these limits could be attained, even in principle,
is a question whose answer will have to await a unified theory
of quantum gravity.  (See Box 2)

The Schwarzschild radius of a 1 kilogram
computer is $R_S=2Gm/c^2=1.485 \times 10^{-27}$ meters.  The entropy
of a black hole is Boltzmann's constant times its area divided by 4,
as measured in Planck units.  Accordingly, the amount of information
that can be stored in a black hole is $I=4\pi G m^2/\ln 2 \hbar c
= 4 \pi m^2/ \ln 2 m_P^2$.
The amount of information that can be stored by the 1 kilogram
computer in the black-hole limit is $3.827 \times 10^{16}$ bits.
A computer compressed to the size of a black hole 
can perform $5.4258 \times 10^{50}$ 
operations per second, the same as the 1 liter computer.

In a computer that has been compressed to its Schwarzschild radius,
the energy per bit is $E/I=mc^2/I=\ln2 \hbar c^3/4\pi m G =
\ln 2 k_B T/2$, where $T=(\partial S/\partial E)^{-1} =
\hbar c/4\pi k_B R_S $ is the temperature of 
the Hawking radiation emitted by the hole.  As a result,
the time it takes to flip a bit on average is $t_{\rm flip}
= \pi\hbar I/2E = \pi^2 R_S/c\ln 2$.  In other words, according
to a distant observer, the amount of time it takes to flip a bit,
$t_{\rm flip}$, is on the same order as the amount of time 
$t_{\rm com}=\pi R_S/c$ it takes to communicate from one side 
of the hole to the other by going around the horizon: 
$t_{\rm com}/t_{\rm flip}=\ln 2/\pi$ .  In contrast to 
computation at lesser densities, which is highly parallel as noted
above, computation at the horizon of a black 
hole is highly serial: every bit is essentially connected to every
other bit over the course of a single logic operation.
As noted above, the serial nature of computation at the black-hole
limit can be deduced from the fact that black holes attain
the Beckenstein bound$^{20-22}$, $k_B RE/\hbar c S = 1/2\pi$.  

\bigskip\noindent{\bf (Box 2: Can a black hole compute?} 

No information can escape from a classical black hole:
what goes in does not come out.  The quantum mechanical picture
of a black hole is different, however.  First of all, 
black holes are not quite black: they radiate at the Hawking temperature.
$T$ given above.  In addition, 
the well-known statement that `a black hole has no hair'---i.e.,
from a distance all black holes with the same charge and angular
momentum look essentially alike  ---  is now known to be not always 
true$^{91-93}$.  Finally, recent work in string theory$^{94-96}$ 
suggests that black holes do not actually destroy the
information about how they were formed, but instead
process it and emit the processed information as part of
the Hawking radition as they evaporate: what does in
does come out, but in an altered form.  

If the latter picture
is correct, then black holes could in principle be 
`programmed': one forms a black hole whose initial conditions
encode the information to be processed, lets that information 
be processed by the Planckian dynamics at the hole's horizon, 
and gets out the answer to the computation by examining the 
correlations in the Hawking radiation emitted when the hole evaporates.
Despite our lack of knowledge of the precise details of
what happens when a black hole forms and evaporates (a full account
must await a more exact treatment using whatever theory of quantum
gravity and matter turns out to be the correct one), we can
still provide a rough estimate how much information is processed during
this computation.  Using Page's results on the rate of
evaporation of a black hole$^{88}$, we obtain a lifetime for the
hole $t_{\rm life} = G^2 m^3/3 C \hbar c^4$, where $C$ is
a constant that is depends on the number of species of
particles with a mass less than $k_B T$, where $T$ is the
temperature of the hole.  For $O(10^1-10^2)$ such species,
$C$ is on the order of $10^{-3}-10^{-2}$, leading to a 
lifetime for a 1 kilogram black hole of $\approx 10^{-19}$
seconds, during which time the hole can perform $\approx
10^{32}$ operations on its $\approx 10^{16}$ bits.
As the actual number of effectively massless particles
at the Hawking temperature of a one-kilogram black hole
is likely to be considerably larger than $10^2$, this number should
be regarded as an upper bound on the actual number of
operations that could be performed by the hole.
Interestingly, although this hypothetical computation is performed
at ultra-high densities and speeds, the total 
number of bits available to be processed is not far 
from the number available to current computers operating 
in more familiar surroundings. {\bf )} 

\bigskip\noindent{\bf 4. Constructing ultimate computers}

Throughout this entire discussion of the physical limits
to computation, no mention has been made of how to construct
a computer that operates at those limits.  In fact, contemporary
quantum `microcomputers' such as those constructed using nuclear
magnetic resonance$^{76-80}$  do indeed operate at the limits
of speed and memory space described above.  Information is stored
on nuclear spins, with one spin registering one bit. 
The time it takes a bit to flip from a state $|\uparrow\rangle$
to an orthogonal state $|\downarrow\rangle$ is
given by $\pi \hbar/2\mu B = \pi\hbar/ 2 E $, where 
$\mu$ is the spin's magnetic moment, $B$ is the magnetic field,
and $E=\mu B$ is the average energy of interaction between the
spin and the magnetic field.
To perform a quantum logic operation between two 
spins takes a time $\pi\hbar/2E_\gamma$, where $E_\gamma$ is the 
energy of interaction between the two spins.  

Although NMR quantum computers already operate at the limits
to computation set by physics, they are nonetheless much slower
and process much less information than the ultimate laptop 
described above.  This is because their energy is largely 
locked up in mass, thereby limiting both their
speed and their memory.  Unlocking this energy is of course possible,
as a thermonuclear explosion indicates.  Controlling such an 
`unlocked' system is another question, however.  In discussing 
the computational power of physical systems in which all energy
is put to use, we assumed that such control is possible in principle,
although it is certainly not possible in current practice.
All current designs for quantum computers operate at low energy
levels and temperatures, exactly so that precise control can
be exerted on their parts.  

As the discussion of error correction above indicates, 
the rate at which errors can be detected and rejected 
to the environment by error correction routines puts a 
fundamental limit on the rate at which errors can be committed.
Suppose that each logical operation performed by the ultimate
computer has a probability $\epsilon$ of being erroneous.
The total number of errors committed by the ultimate computer 
per second is then $2\epsilon E/\pi\hbar$.  The maximum
rate at which information can be rejected to the environment
is, up to a geometric factor, $\ln2 c S/R$ (all bits in the computer
moving outward at the speed of light).  Accordingly,
the maximum error rate that the ultimate computer can
tolerate is $\epsilon \leq \pi \ln2 \hbar c S/2ER =
2 t_{\rm flip}/t_{\rm com}$.  That is, the maximum error
rate that can be tolerated by the ultimate computer is the
inverse of its degree of parallelization. 
 
Suppose that control of highly 
energetic systems were to become possible.  Then how might these 
systems be made to compute?  
As an example of a `computation' that might be performed
at extreme conditions, consider a heavy-ion collision that
takes place in heavy-ion collider at Brookhaven$^{97}$.  If one
collides 100 on 100 nucleons at 200 GeV per nucleon, the
operation time is $\pi\hbar/2E \approx 10^{-29}$ seconds.
The maximum entropy can be estimated to be approximately
to be $4$ per relativistic pion (to within a factor of less
than 2 associated with the overall production rate
per mesons) of which there are approximately $10^4$ per
central collision in which only a few tens of nucleons
are spectators.  Accordingly, the total amount of memory
space available is $S/k_B \ln 2 \approx 10^4 - 10^5$ bits.
The collision time is short: in the center of mass frame
the two nuclei are Lorentz contracted to $D/\gamma$ where
$D= 12-13$ fermi and $\gamma=100$, giving a total collision
time of $\approx 10^{-25}$ seconds.  During the collision,
then, there is time to perform approximately $10^4$
operations on $10^4$ bits --- a relatively simple computation.
(The fact that only one operation per bit is performed suggests
that there is insufficient time to reach thermal equilibrium,
an observation that is confirmed by detailed simulations.) 
The heavy ion system could be programmed by manipulating and
preparing the initial momenta and internal nuclear states of the ions.  
Of course, one does not expect to be able do word processing
on such a `computer.'  Rather one expects to uncover basic knowledge
about nuclear collisions and quark-gluon plasmas: in the
words of Heinz Pagels, the plasma `computes itself.'$^{98}$ 

At the greater extremes of a black hole computer, we assumed
that whatever theory (string theory, M theory?) turns out to be
the correct theory of quantum matter and gravity, it is possible
to prepare initial states of such systems that causes their natural
time evolution to carry out a computation.  What assurance do we have
that such preparations exist, even in principle?

Physical systems that can be programmed to perform arbitrary
digital computations are called computationally universal.
Although computational universality might at first seem to be a
stringent demand on a physical system, a wide variety of physical
systems --- ranging from nearest neighbor Ising
models$^{52}$ to quantum electrodynamics$^{84}$ and 
conformal field theories$^{86}$ --- 
are known to be computationally universal$^{51-53, 55-65}$.  Indeed,
computational universality seems to be the rule rather than the
exception.  Essentially any quantum system that admits controllable
nonlinear interactions can be shown to be computationally universal$^{60-61}$.
For example, the ordinary electrostatic interaction between two
charged particles can be used to perform universal quantum logic
operations between two quantum bits.  A bit is registered by the
presence or absence of a particle in a mode.
The strength of the interaction between the particles,
$e^2/r$, determines the amount of time 
$ t_{flip}=\pi \hbar r/2 e^2$ it takes
to perform a quantum logic operation such as a Controlled-$NOT$
on the two particles. 
Interestingly, the time it takes to perform such an operation
divided by the amount of time it takes to send a signal at
the speed of light between the bits $t_{\rm com} = r/c$
is a universal constant, $t_{flip}/t_{\rm com} = 
\pi\hbar c/ 2 e^2 = \pi/2 \alpha$, where $\alpha
= e^2/\hbar c \approx 1/137$ is the fine structure constant.
This example shows the degree to which the laws of physics
and the limits to computation are entwined.

In addition to the theoretical evidence that most systems are
computationally universal, the computer on which I am writing this 
article provides
strong experimental evidence that whatever the correct underlying
theory of physics is, it supports universal computation.  Whether
or not it is possible to make computation take place in the extreme
regimes envisaged in this paper is an open question.  The answer
to this question lies in future technological development, which
is difficult to predict.  If, as
seems highly unlikely, it is possible to extrapolate the 
exponential progress of Moore's law into the future, then it
will only take two hundred and fifty years to make up the forty orders
of magnitude in performance between current computers that perform
$10^{10}$ operations per second on $10^{10}$ bits and our 
one kilogram ultimate laptop that performs $10^{51}$ 
operations per second on $10^{31}$ bits.

\vfill
\eject

\centerline{\bf References}

\bigskip
\noindent 

\item{1.} Maxwell, J.C., {\it Theory of Heat}, (Appleton, London),
1871.

\item{2.} Smoluchowski, M. von, {\it Z. Physik}, {\bf 13}, 1069 (1912). 

\item{3.} Szilard, L., {\it Z. Physik}
{\bf 53}, 840-856 (1929).

\item{4.} Brillouin, L., `Science and Information Theory,'
(Academic Press, New York), 1953.

\item{5.} Landauer, R., {\it IBM J. Res. Develop.} {\bf 5},
183 (1961).

\item{6.} Keyes, R.W., Landauer, R., {\it IBM Journal of Research
and Development}, {\bf 14}, no. 2, 152-157 (1970).

\item{7.} Landauer, R., {\it Nature}, {\bf 335}, 779-784 (1988).

\item{8.} Landauer, R.,  {\it Physics Today}, May 1991, 23-29 (1991).

\item{9.} Landauer, R.,  {\it Phys. Lett. A}, {\bf 217}, 188-193 (1996).

\item{10.} von Neumann, J., `Theory of Self-Reproducing Automata,'
Lecture 3 (University of Illinois Press, Urbana), 1966.

\item{11.} Lebedev, D.S., Levitin, L.B., {\it Information and Control}
{\bf 9}, 1-22 (1966).

\item{12.} Levitin, L.B., in  `Proceedings of the 3rd International
Symposium on Radio Electronics,' part 3, 1-15 (Varna, Bulgaria),
1970. 

\item{13.} L.B. Levitin,  {\it Int. J. Theor. Phys.} {\bf 21}, 299-309
(1982).

\item{14.} Levitin, L.B. {\it Physica D} {\bf 120}, 162-167 (1998).

\item{15.} Margolus, N., Levitin, L.B., in {\it PhysComp96}, T. Toffoli,
M. Biafore, J. Leao, eds. (NECSI, Boston) 1996.

\item{16.} Margolus, N., Levitin, L.B.,  {\it Physica D} {\bf 120}, 
188-195 (1998). 

\item{17.} Bremermann, H.J., in `Self-Organizing Systems,'
M.C. Yovits, G.T. Jacobi, and G.D. Goldstein, eds.
(Spartan Books, Washington, D.C.), 1962. 

\item{18.} Bremermann, H.J., in `Proceedings of the Fifth
Berkeley Symposium on Mathematical Statistics and Probability,'
(University of California Press, Berkeley), 1967.

\item{19.} Bremermann, H.J., {\it Int. J. Theor. Phys.} {\bf 21}, 203-217 
(1982).

\item{20.} Bekenstein, J.D., {\it Phys. Rev. D.} {\bf 23}, 287 (1981).

\item{21.} Bekenstein, J.D., {\it Phys. Rev. Letters\/} {\bf 46}, 623 (1981).

\item{22.} Bekenstein, J.D., {\it Phys. Rev. D.} {\bf 30}, 1669-1679 (1984).

\item{23.} Aharonov, Y., Bohm, D., {\it Phys. Rev.} {\bf 122},
1649 (1961).

\item{24.} Aharonov, Y., Bohm, D., {\it Phys. Rev. B.} {\bf 134} {\bf 122},
1417 (1964).

\item{25.} Anandan, J., Aharonov, Y. {\it Phys. Rev. Lett.} {\bf
65}, 1697 (1990). 

\item{26.} Peres, A., {\it Quantum Theory:
Concepts and Methods}, (Kluwer, Hingham) 1995.  

\item{27.} Lecerf, Y. {\it Comptes Rendus} {\bf 257}, 2597
(1963).

\item{28.} Bennett, C.H., {\it IBM J. Res. Develop.} {\bf 17},
525 (1973) 

\item{29.} Bennett, C.H., {\it Int. J. Theor. Phys.} {\bf 21},
905 (1982).  

\item{30.} Bennett, C.H., {\it Sci. Am.} {\bf 257},
108 (1987).

\item{31.} Fredkin, E., Toffoli, T., {\it Int. J. Theor. Phys.} {\bf
21}, 219 (1982).

\item{32.} K.K. Likharev {\it Int. J. Theor. Phys.}, {\bf 21}, 311-325 
(1982).

\item{33.} Seitz, C.L. {\it et al.}, in {\it Proceedings of the
1985 Chapel Hill Conference on VLSI}, (Computer Science Press) 1985.

\item{34.} Merkle, R.C., `Reversible Electronic Logic Using
Switches,' submitted to {\it Nanotechnology}, 1992.

\item{35.} Younis, S.G., Knight, T.F., in {\it Proceedings
of the 1993 Symposium on Integrated Systems}, Seattle, Washington,
(MIT Press, Cambridge) 1993.

\item{36.} Lloyd, S., Pagels, H. {\it Ann. Phys.} {\bf 188}, 
186-213, (1988).

\item{37.} Lloyd, S., {\it Phys. Rev. A} {\bf 39}, 5378-5386, (1989).

\item{38.} Zurek, W.H., {\it Nature} {\bf 341}, 119 (1989).

\item{39.} Leff, H.S., Rex, A.F., {\it Maxwell's Demon:
Entropy, Information, Computing,} Princeton University Press,
Princeton, 1990.

\item{40.} Lloyd, S., {\it Phys. Rev. A} {\bf 56}, 3374-3382, (1997).

\item{41.} Benioff, P., {\it J. Stat. Phys.} {\bf 22}, 563
(1980). 

\item{42.} Benioff, P., {\it Phys. Rev. Lett.} {\bf 48}, 1581
(1982).

\item{43.} Feynman, R.P., {\it Int. J. Theor.
Phys.} {\bf 21}, 467 (1982).

\item{44.} Feynman, R.P., {\it Optics News} {\bf 11}, 11 (1985);
reprinted in {\it Found. Phys.} {\bf 16}, 507 (1986). 

\item{45.} Zurek, W.H., {\it Phys. Rev. Lett.} {\bf 53}, 391
(1984).

\item{46.} Peres, A., {\it Phys. Rev.} {\bf A 32}, 3266 (1985).

\item{47.} Deutsch, D., {\it Proc. Roy. Soc. Lond.} {\bf A 400},
97 (1985). 

\item{48.} Margolus, N., {\it Ann. N.Y. Acad. Sci.} {\bf 480},
487 (1986). 

\item{49.} Deutsch, D., {\it Proc. Roy. Soc. Lond.} {\bf A 425}, 73
(1989).

\item{50.} Margolus, N., {\it Complexity, Entropy, and the Physics of
Information, Santa Fe Institute Studies in the Sciences of
Complexity} {\bf VIII}, W.H. Zurek, ed., (Addison Wesley, Redwood City,
1991) pp. 273-288.

\item{51.} Lloyd, S., {\it Phys. Rev. Lett.} {\bf 71}, 943 (1993).

\item{52.} Lloyd, S., {\it Science} {\bf 261}, 1569-1571 (1993).

\item{53.} Lloyd, S., {\it J. Mod. Opt.} {\bf 41}, 2503 (1994).

\item{54.} Shor, P., in 
{\it Proceedings of the 35th Annual Symposium on Foundations
of Computer Science}, S. Goldwasser, Ed., IEEE Computer
Society, Los Alamitos, CA, 1994, pp. 124-134.

\item{55.} Lloyd, S., {\it Sci. Am.} {\bf 273}, 140 (1995).

\item{56.} Divincenzo, D., {\it Science} {\bf 270}, 255 (1995).
 
\item{57.} DiVincenzo, D.P., {\it Phys. Rev. A} {\bf 51},
1015-1022 (1995).

\item{58.} Sleator, T., Weinfurter, H., {\it Phys. Rev. Lett.} 
{\bf 74}, 4087-4090 (1995).

\item{59.} Barenco, A., Bennett, C.H., Cleve, R., DiVincenzo, D.P.,
Margolus, N., Shor, P., Sleator, T., Smolin, J.A., Weinfurter, H.,
{\it Phys. Rev. A} {\bf 52}, 3457-3467 (1995).

\item{60.} Lloyd, S., {\it Phys.\ Rev.\ Lett.\/}, {\bf 75},
346-349 (1995).

\item{61.} Deutsch, D., Barenco, A., Ekert, A., {\it Proc.\
Roy.\ Soc.\ A} {\bf 449}, 669-677 (1995).
 
\item{62.} Cirac, J.I., Zoller, P. {\it Phys. Rev. Lett.} {\bf 74}, 
4091-4094 (1995).

\item{63.} Pellizzari, T., Gardiner, S.A., Cirac, J.I., Zoller, P.,
{\it Phys. Rev. Lett.} {\bf 75}, 3788-3791 (1995).

\item{64.} Turchette, Q.A., Hood, C.J., Lange, W., Mabuchi, H.,
Kimble, H.J., {\it Phys. Rev. Lett.} {\bf 75}, 4710-4713 (1995).

\item{65.} Monroe, C., Meekhof, D.M., King, B.E., Itano, W.M., 
Wineland, D.J., {\it Phys. Rev. Lett.} {\bf 75}, 4714-4717 (1995).

\item{66.} Grover, L.K., in {\it Proceedings of the 28th Annual ACM Symposium
on the Theory of Computing}, ACM Press, New York, 1996, pp. 212-218.

\item{67.} Lloyd, S.,  {\it Science} {\bf 273}, 1073-1078 (1996).

\item{68.} Zalka, C., {\it Proc. R. Soc. Lond A. Mat.} {\bf 454}, 
313-322 (1998).

\item{69.} Shor, P.W., {\it Phys. Rev. A} {\bf 52}, R2493-R2496 (1995).

\item{70.} Steane, A.M., {\it Phys. Rev. Lett.}
{\bf 77}, 793-797 (1996).

\item{71.} Laflamme, R., Miquel, C., Paz, J.P., 
Zurek, W.H., {\it Phys. Rev. Lett.} {\bf 77}, 198-201 (1996).

\item{72.} DiVincenzo, D.P., Shor, P.W., {\it
Phys.\  Rev.\ Lett.\/} {\bf 77}, 3260-3263 (1996).

\item{73.} Shor, P., in
{\it Proceedings of the 37th Annual Symposium on the Foundations
of Computer Science,} IEEE Computer Society Press, Los Alamitos,
1996, pp. 56-65.

\item{74.} Preskill, J., {\it Proc. Roy. Soc. Lond. Ser. A} {\bf 454},
385 (1998).

\item{75.} Laflamme, R., Knill, M., Zurek, W.H., {\it
Science}  {\bf 279}, 342 (1998).   

\item{76.} Cory, D.G., Fahmy, A.F., Havel, T.F., 
in {\it PhysComp96, Proceedings of the
Fourth Workshop on Physics and Computation}, T. Toffoli, M. Biafore,
J. Le\~ao, eds., (New England Complex Systems Institute, Boston) 1996.

\item{77.} Gershenfeld, N.A., Chuang, I.L. 
{\it Science} {\bf 275}, 350-356
(1997).

\item{78.} Chuang, I.L., Vandersypen, L.M.K.,
Zhou, X., Leung, D.W., Lloyd, S., {\it Nature} {\bf 393},
143-146 (1998) May, 1998.

\item{79.} Jones, J.A., Mosca, M., Hansen, R.H., {\it Nature}
{\bf 393}, 344-346 (1998). 

\item{80.} Chuang, I.L., Gershenfeld, N., Kubinec, M., {\it Phys. Rev.
Lett.} {\bf 80}, 3408-3411 (1998).

\item{81.} Kane, B., {\it Nature} {\bf 393}, 133 (1998).

\item{82.} Nakamura, Y., Pashkin, Yu.A., Tsai, J.S., {\it Nature}
{\bf 398}, 305 (1999)

\item{83.} Mooij, J.E., Orlando, T.P., Levitov, L., Tian, L.,
van der Wal, C.H., Lloyd, S., {\it Science} {\bf 285}, 1036-1039
(1999).

\item{84.} Lloyd, S., Braunstein, S., {\it Phys. Rev. Lett.}
{\bf 82}, 1784-1787 (1999).

\item{85.} Abrams, D., Lloyd, S.,  {\it Phys. Rev. Lett.} {\bf 81},
3992-3995, (1998).
 
\item{86.} Freedman, M., unpublished.

\item{87.} Zel'dovich, Ya. B., Novikov, I.D., {\it Relativistic
Astrophysics,} (Univ. of Chicago Press, Chicago), 1971.

\item{88.} Page, D.N., {\it Phys. Rev. D} {\bf 13}, 198 (1976).

\item{89.} K.S. Thorne, R.H. Price, D.A. Macdonald, {\it Black Holes:
The Membrane Paradigm,} (Yale, New Haven) 1986.  In particular,
chapter VIII, `The Thermal Atmosphere of a Black Hole,' by
K.S. Thorne, W.H. Zurek, and R.H. Price.

\item{90.} Novikov, I.D., Frolov, V.P., {\it Black Holes,} 
(Springer-Verlag, Berlin) 1986.

\item{91.} Coleman, S.,
Preskill, J., Wilczek, F., {\it Phys. Rev. Lett.} {\bf 67}
1975-1978 (1991). 

\item{92.} Preskill, J., {\it Phys. Scr. T } {\bf 36}, 
258-264 (1991).  

\item{93.} Fiola, T.M., Preskill, J., Strominger A., 
{\it Phys. Rev. D} {\bf 50}, 3987-4014 (1994).  

\item{94.} Susskind, L., Uglum, J., {\it Phys. Rev. D}
{\bf 50}, 2700 (1994). 

\item{95.} Strominger A., Vafa, C., {\it Phys. Lett. B} 
{\bf 37}, 99 (1996). 

\item{96.} Das, S.R., Mathur, S.D., {\it Nucl. Phys. B} {\bf 478}, 561-576
(1996).

\item{97.} Sidney H. Kahana, private communication. 

\item{98.} Pagels, H., {\it The Cosmic Code: quantum physics as the
language of nature}, (Simon and Schuster, New York) 1982.

\vfill\eject

\bigskip
\noindent{\bf Figure 1: The Ultimate Laptop}

The `ultimate laptop' is a computer with a mass of one kilogram
and a volume of one liter, operating at the fundamental limits
of speed and memory capacity fixed by physics.  The ultimate
laptop performs $2mc^2/ \pi \hbar = 5.4258 \times 10^{50}$
logical operations per second on $\approx 10^{31}$ bits.
Although its computational machinery is in fact in a highly
specified physical state with zero entropy,
while it performs a computation that uses all its resources
of energy and memory space it appears to an outside observer
to be in a thermal state at $\approx 10^9$ degrees Kelvin.
The ultimate laptop looks like a small piece of the Big Bang.

\bigskip
\noindent{\bf Figure 2: Computing at the Black-Hole Limit}

The rate at which the components of a computer can communicate
is limited by the speed of light.  In the ultimate laptop,
each bit can flip $\approx 10^{19}$ times per second,
while the time to communicate from one side of the one
liter computer to the other is on the order of $10^9$
seconds: the ultimate laptop is highly parallel.
The computation can be speeded up and made more serial
by compressing the computer.  But no computer can be
compressed to smaller than its Schwarzschild radius without
becoming a black hole.  A one-kilogram computer that has been 
compressed to the black hole limit of
$R_S=2Gm/c^2=1.485 \times 10^{-27}$ meters can perform 
$ 5.4258 \times 10^{50}$ operations per second on 
its $I=4\pi G m^2/\ln 2 \hbar c = 3.827 \times 10^{16}$ bits.  
At the black-hole limit,
computation is fully serial: the time it takes to flip
a bit and the time it takes a signal to communicate
around the horizon of the hole are the same.  

\vfill\eject\end

\item{99.} In this paper we have explored only one aspect of
how a black hole could be used for computation. Jim Hartle
(private communication) has suggested some alternative methods:

\item{(i)} Set up a foundation to perform a very long computation to
which you wish to know the answer.  Then retire to near the
horizon of the hole.  Since your clock is redshifted with
respect to the foundation's computer's clock, when you emerge
from the hole a few minutes later by your clock,
many years may have passed by the external clock (i.e.,
the black hole can be used as a time machine to travel
into the future).  As long as you have left the foundation
in reliable hands so that it actually performed the desired
computation, you can now get the answer.

\item{(ii)} If you really want the answer to the computation
sufficiently badly, you can use the following feature of certain
black holes.  In some black holes there are time-like
curves passing through the horizon that intersect all other 
time-like curves that pass the horizon at later external time
(M. Simpson and R. Penrose, {\it Int. J. Theor. Phys.} {\bf 7},
183-197 (1973); S. Chandrasekar and J.B. Hartle, {\it
Proc. R. Soc. Lond. Ser. A} {\bf 384}, 301-315 (1982)): an observer
travelling along such a curve will obtain during a finite interval
of his time all information that subsequently falls into the
hole.  In this case, the foundation you set up can take an
arbitrarily long time to perform the computation.  Of course,  
you will be stuck inside the hole when you get the answer, which 
you will have to extract from the highly blue-shifted radiation 
falling into the hole.  (In addition, the fact that you
are obtaining a potentially infinite amount of information in
a finite time suggests that travelling along such trajectories may 
be hazardous.)